\documentclass[fleqn,usenatbib]{mnras}

\usepackage{newtxtext,newtxmath}

\usepackage[T1]{fontenc}

\DeclareRobustCommand{\VAN}[3]{#2}
\let\VANthebibliography\thebibliography
\def\thebibliography{\DeclareRobustCommand{\VAN}[3]{##3}\VANthebibliography}

\usepackage{graphicx}	
\usepackage{amsmath}	

\title[III Zw 2]{Outflows in the Radio-Intermediate Quasar III Zw 2: A Polarization Study with the EVLA \& uGMRT}
\author[Silpa et al.]{Silpa S.,$^{1}$\thanks{E-mail: silpa@ncra.tifr.res.in}
P. Kharb,$^{1}$
C. M. Harrison,$^{2}$
L. C. Ho,$^{3,4}$
M. E. Jarvis,$^{5,6,7}$
C. H. Ishwara-Chandra,$^{1}$
B. Sebastian$^{8}$\\
$^{1}$National Centre for Radio Astrophysics $-$ Tata Institute of Fundamental Research, S. P. Pune University Campus, Ganeshkhind, Pune 411007, India \\
$^{2}$School of Mathematics, Statistics and Physics, Newcastle University, Newcastle upon Tyne, NE1 7RU, UK\\
$^{3}$Kavli Institute for Astronomy and Astrophysics, Peking University, Beijing 100871, China\\
$^{4}$Department of Astronomy, School of Physics, Peking University, Beijing 100871, China\\
$^{5}$Max-Planck Institut f\"ur Astrophysik, Karl-Schwarzschild-Str. 1, 85748 Garching, Germany\\
$^{6}$European Southern Observatory, Karl-Schwarzschild-Str. 2, 85748 Garching, Germany\\
$^{7}$Ludwig Maximilian Universit\"at, Professor-Huber-Platz 2, 80539 Munich, Germany\\
$^{8}$Department of Physics and Astronomy, Purdue University, 525 Northwestern Avenue, West Lafayette, IN 47907, USA\\
}

\pubyear{2021}

\begin{document}
\label{firstpage}
\pagerange{\pageref{firstpage}--\pageref{lastpage}}
\maketitle

\begin{abstract}
We present results from a polarization study of the radio-intermediate quasar, III~Zw~2, at a redshift of 0.089, with the upgraded Giant Metrewave Radio Telescope (uGMRT) at 685~MHz and the Karl G. Jansky Very Large Array (VLA) at 5 and 34 GHz. We detect a kpc-scale outflow, exhibiting transverse magnetic (B-) fields. The curved jet terminates in a bow-shock-like radio structure with inferred B-fields aligned with the lobe edges. We suggest that the radio outflow in III~Zw~2 is a combination of a collimated jet along with a wind-like component. This ``wind'' component could be a magnetized accretion disk wind or the outer layers of a broadened jet or a combination of both. The current data cannot differentiate between these possibilities. We also detect kpc-scale lobe emission that is misaligned with the primary lobes in the uGMRT images. The spectral indices and the electron lifetimes in the misaligned lobe are similar to the primary lobe, suggesting that the misaligned lobe is not a relic. We propose that changing spectral states of the accretion disk, and the subsequent intermittent behaviour of the outflow, along with the close interplay between the jet and ``wind'' could explain the radio-intermediate nature of III Zw 2. Our study shows that radio-intermediate quasars are promising sources for understanding the role of jets and winds in galaxy evolution and demonstrates the power of radio polarization studies towards achieving this.
\end{abstract}

\begin{keywords}
(galaxies:) quasars: individual: III Zw 2 -- radio continuum: general -- techniques: polarimetric 
\end{keywords}



\section{Introduction}
\label{sec:intro}
Seyfert galaxies are a subclass of active galactic nuclei (AGN). AGN are powered by accretion of matter on to supermassive black holes (SMBH). Type 1 Seyferts are those that exhibit the presence of both broad and narrow emission lines in their optical spectra, while Type 2 exhibit only narrow emission lines; these two types have been explained to be due to orientation and obscuration effects \citep{Antonucci93}. Seyfert galaxies are typically classified as radio-quiet (RQ) AGN, based on their radio-loudness parameter (R) which is the ratio of 5~GHz radio flux density to optical B-band flux density being $\le10$ \citep{Kellermann89}. However, majority of the type 1 Seyfert galaxies move into the radio-loud (RL) AGN category when only the nuclear contributions in both radio and optical flux densities are considered \citep{HoPeng01,Kharb14}. The origin of radio outflows in Seyfert galaxies is poorly understood \citep{Panessa19}. AGN jets/winds and starburst-driven winds are considered as potential contributors to the radio emission observed in Seyfert galaxies \citep{Baum93,Colbert96,Gallimore06}. Radio polarimetric observations can in principle help to distinguish between these primary contenders based on the differences in their degrees of polarization, magnetic (B-) field structures and rotation measures (RM) \citep[e.g.,][]{Sebastian19b,Sebastian20}.

It has long been suggested that the radio-intermediate (RI) quasars \citep[i.e. those with 10 $<$ R $<$ 250;][]{Falcke96} are relativistically boosted counterparts of RQ quasars, III~Zw~2 \citep[with R = 200;][]{Kellermann94} being a classic example \citep{Miller93, Falcke96}. In the light of this hypothesis, studying outflows in RI quasars can provide better insights into the origin of radio emission in RQ quasars and on how they compare with the properties of the better studied class of RL quasars. 

Apart from radiation, accretion of matter onto SMBHs can drive powerful outflows, such as jets and/or winds, which transport matter and energy from the nucleus to the surrounding medium. These outflows can affect star-formation, chemical composition of the ambient medium and cooling flows at the galaxy clusters cores \citep[e.g.,][]{Fabian12}. They play an important role in the co-evolution of SMBHs and their host galaxies, \citep[the well-known M-$\sigma$ relation;][]{FerrareseMerritt00}, via the ``AGN feedback'' processes \citep[e.g.,][]{Mullaney13,Jarvis19,Zakamska14, Smith20}. However, there are several gaps in our understanding about the different modes of these outflows, such as jets and winds, and how each influence their host galaxies. This problem could be better addressed if studied in systems where more than one outflow mode exists. \citet{Mehdipour19} have suggested that two modes of outflow, i.e. a jet and a wind, could co-exist in AGN of intermediate radio-loudness (see their Figure 4). Therefore, RI AGN may be promising sources for understanding how AGN influence their host galaxies and the role of jets/winds in galaxy evolution.

While jets are generally understood to be produced by the electromagnetic extraction of rotational energy from spinning black holes \citep{BlandfordZnajek77}, winds may originate either from the accretion disk or the torus or the broad line region and may be driven either thermally \citep{Begelman83,Woods96,KrolikKriss01,Mizumoto19}, or radiatively \citep{ProgaKallman04,ZakamskaGreene14,Nims15}, or magnetically \citep{BlandfordPayne82, KoniglKartje94, Everett05, Fukumura10}. Another model for nuclear ``wind'' is proposed by \citet{Mukherjee18} in their numerical simulations, where they demonstrate how the strong interaction between an inclined relativistic jet and the dense interstellar medium (ISM) can cause the jet to deflect, decelerate, and break out into a sub-relativistic outflow reminiscent of an AGN ``wind'' \citep{RupkeVeilleux11} or a nuclear starburst \citep{Veilleux05}, along the minor axis of the galaxy following a path of least resistance. Such a ``wind'' creates a spherical bubble that clears the ambient medium as it rises. 

Therefore, one cannot fully differentiate between a magnetized accretion disk wind and the outer layers of a broadened jet (like a jet sheath). To fully differentiate between these, one will need to take into account the emission-line gas structure and kinematics along with the polarized radio emission, as well as realistic magnetohydrodynamic (MHD) jet simulations comprising of these multiple components. Such an analysis is beyond the scope of the present work, but can be aimed for in the future. Therefore, in this paper, we use ``wind'' generally to refer to either a magnetized accretion disk wind or the outer layers of a broadened jet, or a combination of both.

III~Zw~2 has been identified as a Seyfert 1 galaxy in the literature \citep{Arp68, Hutchings83}. III~Zw~2 was in fact the first Seyfert galaxy where a superluminal radio jet was discovered \citep{Brunthaler00}. It is a member of the Palomar Green quasar sample \citep[PG 0007+106;][]{SchmidtGreen83}. The $\it{Fermi}$-LAT detection of two distinct $\gamma$-ray flares in III~Zw~2 has been recently reported \citep{Liao16}. III~Zw~2 is known to be highly variable at radio \citep{Wright77, Schnopper78, Landau80, Aller85, Falcke99}, IR \citep{LebofskyRieke80, Sembay87}, UV \citep{Chapman85}, optical \citep{Lloyd84, Clements95}, X-ray \citep{KaastraKorte88, Salvi02} and $\gamma$-ray \citep{Liao16} wavelengths. Its radio core flux density has been known to vary by a factor of 20-30 over a span of a few years \citep{Aller85, Falcke99}. 

Early studies of III~Zw~2 suggested a spiral host galaxy \citep{HutchingsCampbell83,Taylor96}. However, what was initially identified as a spiral arm by \citet{Hutchings83} is now accepted to be a star-forming tidal arm indicative of an ongoing galaxy merger \citep{Surace01}. Recent studies using Hubble Space Telescope also suggest an elliptical host galaxy \citep{Veilleux09}. At a redshift of 0.0893, it forms a part of a triple interconnected system of compact galaxies \citep{Zwicky71}.

\begin{table*}
\caption{Observation details}
\label{tab:Table1}
\begin{tabular}{ccccccccc}
\hline
Project ID & Telescope Array & Frequency & Observation &
Resolution & Flux & Phase & Polarized & Unpolarized\\
& Configuration &  & Date &
(arcsec) & Calibrator & Calibrator & Calibrator & Calibrator\\
\hline
DDTC130 & uGMRT & 685 MHz & 2020 Jun 23 & 3.6 & 3C48 & IAU 2330+110 & 3C138 & 3C84 \\
20A-182 & VLA B-array & 5 GHz & 2020 Aug 20 & 1.1 & 3C138 & J0010+1724 & 3C138 & 3C84 \\
17A-027 & VLA C-array & 34 GHz & 2017 Jun 16 & 0.6 & 3C48 & 3C48, 3C84 & 3C48 & 3C84 \\
15A-070 & VLA B-array & 1.5 GHz & 2015 Feb 23 & 3.6 & 3C48 & J2355+4950 & ... & ...\\
\hline
\end{tabular}
\end{table*}

The supermassive black hole in III~Zw~2 has been estimated to be $7.4\times10^8$~M$_\odot$ with an Eddington ratio of $\sim$0.07 \citep{Shangguan18}. A part of the BLR is in a Keplerian disk \citep{Popovic03}. Some of the BLR emission may come from an accretion disk wind \citep{Popovic03}. The disk has been inferred to lie at an inclination of $12\pm5\degr$ \citep{Popovic03}. \citet{Mehdipour19} find a significant inverse correlation between ionized wind column density (N$_\mathrm H$) and the radio-loudness parameter for a sample of 16 Seyfert 1s having R $>10$. Their sample includes III~Zw~2 for which they find N$_\mathrm{H}=(7\pm4)\times10^{20}$~cm$^{-2}$, and an ionized wind outflow velocity of $(-1780\pm670)$~km~s$^{-1}$. The N$_\mathrm{H}$-R relation is explained as a manifestation of a common driving mechanism for both outflow modes, such as a magnetically-driven wind and a jet.

Similar to its total intensity, the radio core polarization (angle) in III~Zw~2 varies dramatically over the time period of years \citep{Aller85, Terasranta92}. Jet precession has been inferred from  very long baseline interferometry (VLBI) studies, radio variability, and the X-ray study \citep{Brunthaler05, Li10, Gonzalez18}. A precessing radio jet interacting with a molecular torus roughly every five years was invoked by \citet{Brunthaler05} to explain their observations of radio outbursts in III~Zw~2. A variability period of about 5.14 yrs was discovered in III~Zw~2 by \citet{Li10} based on the analysis of its radio light curves, which they attributed to the helical motion of the radio jet. \citet{Gonzalez18} found that the source of X-ray reflection in III~Zw~2 changed from being an inner accretion disk to a torus over a period of 11 years. This was explained as a result of a precessing radio jet in III~Zw~2 that may have originated from a precessing accretion disk owing to disc instabilities from an ongoing merger event \citep{Surace01, Liska18}.

In this paper, we present the results from a multi-frequency radio polarimetric study of III~Zw~2, and discuss how these results can explain some of the peculiarities of this source. Throughout this work, we have adopted $\Lambda$CDM cosmology with $H_0$ = 73~km~s$^{-1}$~Mpc$^{-1}$, $\Omega_{m}$ = 0.27 and $\Omega_{v}$ = 0.73. Spectral index $\alpha$ is defined such that flux density at frequency $\nu$, $S_\nu\propto\nu^{\alpha}$.

\section{Observations and Data reduction} 
\label{sec:obs}
We observed III Zw 2 using the upgraded Giant Metrewave Radio Telescope (uGMRT) at 685~MHz (Band 4) on June 23, 2020 and the Karl G. Jansky Very Large Array (VLA) at 5~GHz in the B-array configuration on August 20, 2020. The 34~GHz VLA C-array and 1.5~GHz B-array data were obtained from the VLA data archive. The details of the observations are provided in Table~\ref{tab:Table1}. The uGMRT data were reduced with our $\tt{CASA}$\footnote{Common Astronomy Software Applications; \citet{Shaw07}}-based pipeline that handles total intensity \citep[see][]{Silpa20} as well as polarization data\footnote{Available at https://sites.google.com/view/silpasasikumar/}. Basic calibration and editing for VLA 1.5, 5 and 34 GHz data was carried out using the $\tt{CASA}$ calibration pipeline for VLA data reduction. This was followed by manual polarization calibration on 5 and 34 GHz data, since the 1.5 GHz data had only RR and LL visibilities. We discuss the polarization calibration ahead. 

\subsection{Polarization Calibration}
\label{sec:poln_caln}
Firstly, the model of a polarized calibrator was set manually using the task $\tt{SETJY}$ in $\tt{CASA}$. For this, we provided the model parameters such as the reference frequency, Stokes I flux density at the reference frequency, the spectral index, and the coefficients of the Taylor expansion of fractional polarization and polarization angle as a function of frequency about the reference frequency. The coefficients were estimated by fitting a first-order polynomial to the values of fractional polarization and polarization angle as function of frequency, which were obtained from the NRAO VLA observing guide\footnote{https://science.nrao.edu/facilities/vla/docs/manuals/obsguide/modes/pol}. 

\begin{figure*}
\centerline
{\includegraphics[width=17cm,trim=0 0 50 400]{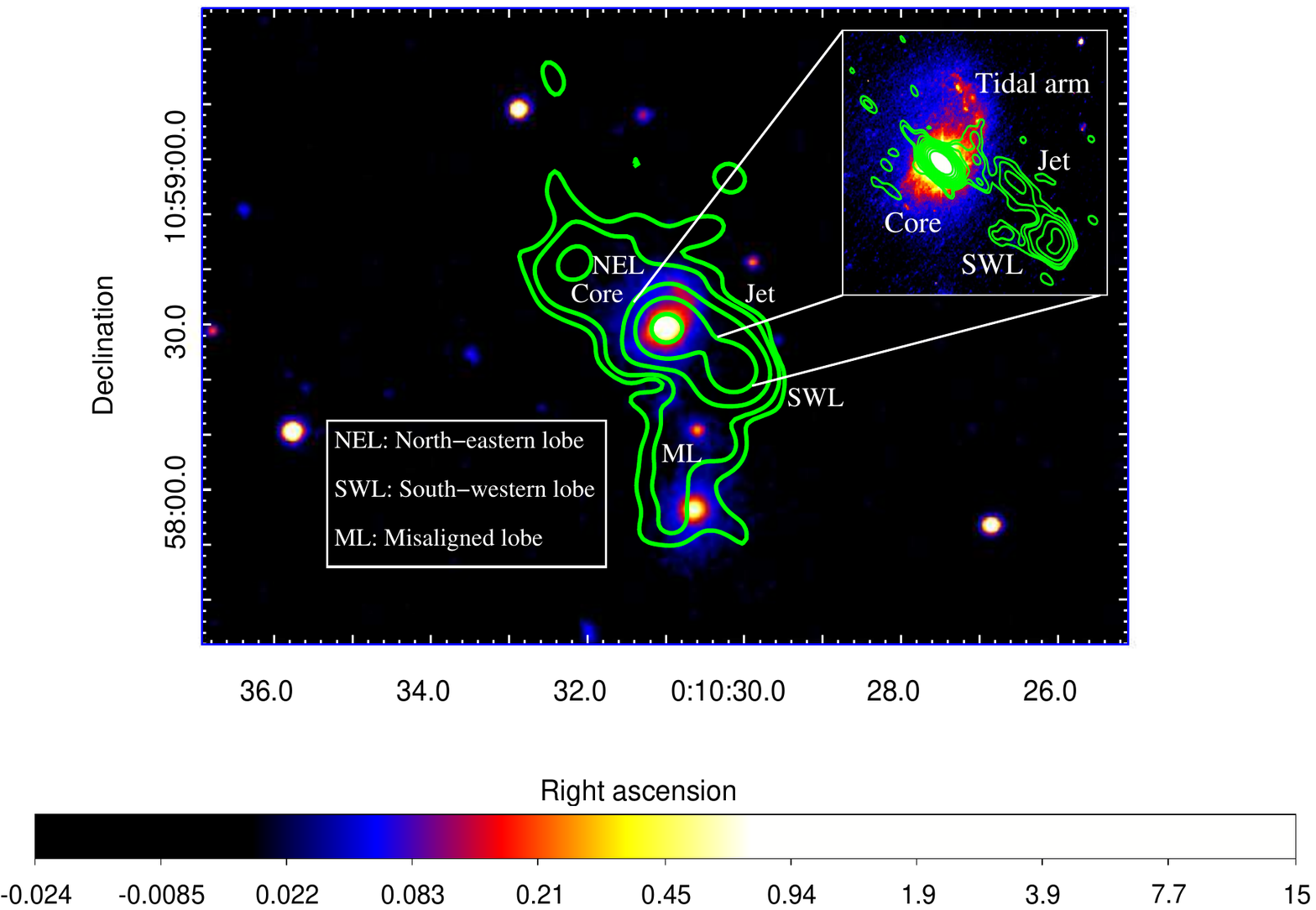}}
\caption{\small uGMRT 685 MHz {\it uv}-tapered total intensity contour image superimposed over SDSS {\it g}-band optical image. The contour levels are (0.40, 0.70, 1.88, 6.58, 25.32) mJy beam$^{-1}$. The color scale extends from $-$0.024 to 15.0 $\times$15 nJy or  
$-0.36$~nJy to 0.22~$\mu$Jy.
The inset on the top right presents VLA 5 GHz total intensity contour image superimposed over the HST WFC3 F547M optical image showing the SW jet and lobe. The enlarged view of the same showing the full radio structure is presented in Figure~\ref{fig:5}.}
\label{fig:1}
\end{figure*}

\begin{figure*}
\centerline
{\includegraphics[width=13cm,trim=0 0 150 400]{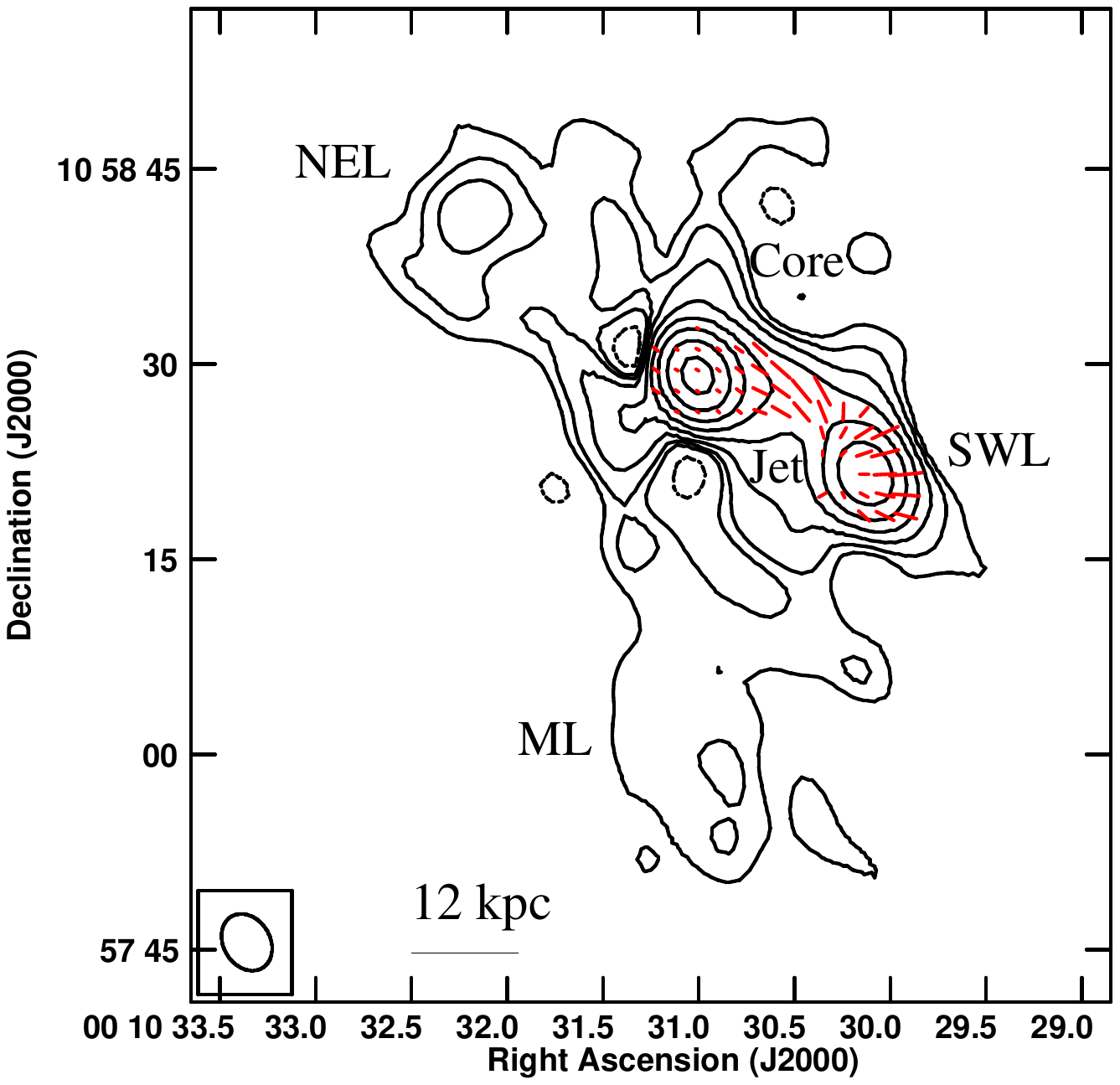}}
\caption{\small uGMRT 685 MHz total intensity contour image with electric fractional polarization vectors superimposed in red. The peak contour flux is 35 mJy beam$^{-1}$ and the contour levels are 0.21 $\times$ (-1, 1, 2, 4, 8, 16, 32, 64, 128, 256, 512) mJy beam$^{-1}$. 2$\arcsec$ length of the vector corresponds to 6.25\% fractional polarization.}
\label{fig:2}
\end{figure*}

Polarization calibration was carried out in three steps: (i) the cross$-$hand (RL, LR) delays, resulting from residual delay difference between R and L signals, were solved using a polarized calibrator that has strong cross-polarization. This was carried out using the task $\tt{GAINCAL}$ with $\tt{gaintype=KCROSS}$ in $\tt{CASA}$. (ii) the instrumental polarization (i.e, the frequency dependent leakage terms or `D-terms'), was solved using either an unpolarized calibrator or a polarized calibrator with good parallactic angle coverage. This corrected for the polarization leakage between the feeds owing to their imperfections and non-orthogonality (for e.g. the R/X$-$polarized feed picks up L/Y-polarized emission, and vice versa). We used the task $\tt{POLCAL}$ in $\tt{CASA}$ to solve for instrumental polarization with $\tt{poltype=Df}$ while using the unpolarized calibrator and with $\tt{poltype=Df+QU}$ while using the polarized calibrator. This task uses the Stokes I, Q, and U values (Q and U being zero for unpolarized calibrator) in the model data to derive the leakage solutions. Also, when $\tt{poltype=Df+QU}$ is used, this task solves for both source polarization and instrumental polarization simultaneously based on their differential rotation with the parallactic angle.

\begin{table*}
\begin{center}
\caption{Summary of radio total intensity and polarization properties}
\label{tab:Table2}
\begin{tabular}{lccccc}
\hline
Project & Component & Total flux & Polarized flux & Fractional & Polarization\\
& & density ($I$, mJy) & density ($P$, mJy) & polarization ($P/I$, \%) & angle ($\chi$, $\degr$) \\
\hline
& Core & 40$\pm$4 & 0.31$\pm$0.03 & 1.0$\pm$0.1 & 57$\pm$3 \\
& Jet & 8.0 & 0.40$\pm$0.04 & 5.4$\pm$0.4 & 41$\pm$3 \\ 
uGMRT 685 MHz & NE lobe & 2.7 & ... & ... & ... \\ & SW lobe & 17.0 & 0.60$\pm$0.04 & 4.8$\pm$0.3 & $-$15$\pm$2 \\
& Misaligned lobe & 8.0 & ... & ... & ... \\ 
\hline
& Core & 126$\pm$12 & 0.12$\pm$0.02 & 0.20$\pm$0.04 & 27$\pm$6 \\ 
VLA 5 GHz & Jet & 1.7 & 0.020$\pm$0.006 & 35$\pm$11 & 26$\pm$8 \\
& NE lobe & 0.5 & ... & ... & ... \\
& SW lobe & 2.9 & 0.22$\pm$0.04 & 16$\pm$3 & 15$\pm$6 \\
\hline
VLA 34 GHz & Core & 539$\pm$53 & 1.7$\pm$0.1 & 0.36$\pm$0.03 & 47$\pm$2 \\ 
\hline
& Core & 88$\pm$8 & ... & ... & ... \\
& Jet & 2.5 & ... & ... & ... \\
VLA 1.5 GHz & NE lobe & 1.8 & ... & ... & ... \\ 
& SW lobe & 10.0 & ... & ... & ... \\
& Misaligned lobe & 5.2 & ... & ... & ... \\
\hline
\end{tabular}
\end{center}
\end{table*}

The unpolarized calibrator 3C84 was used for leakage calibration in all the projects. The average value of the D-term amplitude turned out to be $\sim10\%$ (typically $\le20\%$) for the uGMRT antennas (Project DDTC130). For the VLA antennas, this value turned out to be $\sim7\%$ for Project 20A-182 and $\sim5\%$ for Project 17A-027 (typically $\le10\%$ in both cases). (iii) the frequency-dependent polarization angle was solved using a polarized calibrator with known polarization angle. This corrected for the R-L phase offset which arises from the difference between the right and left gain phases for the reference antenna. This was carried out using the task $\tt{POLCAL}$ in $\tt{CASA}$ with $\tt{poltype= Xf}$. 

The calibration solutions were then applied to the multi-source dataset. The $\tt{CASA}$ task $\tt{SPLIT}$ was used to extract the calibrated visibility data for III Zw 2 from the multi-source dataset while averaging the spectral channels such that the bandwidth smearing effects were negligible. The total intensity or the Stokes I image of III Zw 2 was created using the multiterm-multifrequency synthesis \citep[MT-MFS;][]{RauCornwell11} algorithm of $\tt{TCLEAN}$ task in $\tt{CASA}$. Three rounds of phase-only self-calibration followed by one round of amplitude and phase self-calibration were carried out for the VLA data whereas, four rounds of phase-only self-calibration followed by four rounds of amplitude and phase self-calibration were carried out for the uGMRT data. Stokes Q and U images were created from the last self-calibrated visibility data, using the same input parameters as for the Stokes I image but with lesser number of iterations. The {\it rms} noise of the Stokes~(I,~Q,~U) images is (19, 10, 11)~$\mu$Jy~beam$^{-1}$ for uGMRT at 685~MHz, (66, none, none)~$\mu$Jy~beam$^{-1}$ for VLA at 1.5~GHz, (15, 13, 13)~$\mu$Jy~beam$^{-1}$ for VLA at 5~GHz and (278, 47, 48)~$\mu$Jy~beam$^{-1}$ for VLA at 34~GHz.\footnote{Stokes Q and U images appear to have been overcleaned at 34~GHz compared to Stokes I image. However, potential discrepancies have been accounted for by blanking pixels with Stokes Q or U signal-to-noise ratio less than 3, greater than 10$\degr$ error, and greater than 10\% error, while creating the linear polarization, polarization angle, and fractional polarization images, respectively.}

Linear polarized intensity ($P=\sqrt{Q^2+U^2}$) and polarization angle  ($\chi$ = 0.5 tan$^{-1} (U/Q)$) images were obtained by combining the Stokes Q and U images using the $\tt{AIPS}$\footnote{Astronomical Image Processing System; \citet{Wells85}} task $\tt{COMB}$ with $\tt{opcode=POLC}$ (which corrects for Ricean bias) and $\tt{POLA}$ respectively. Regions with intensity values less than 3 times the {\it rms} noise in the $P$ image and with values greater than 10$\degr$ error in the $\chi$ image were blanked while using $\tt{COMB}$. Fractional polarization ($F=P/I$) images were obtained in $\tt{COMB}$ with $\tt{opcode=DIV}$. Regions with fractional polarization errors $\gtrsim$10\% were blanked in the image. 

\begin{figure*}
\centering{
\includegraphics[width=16cm,trim=0 0 30 420]{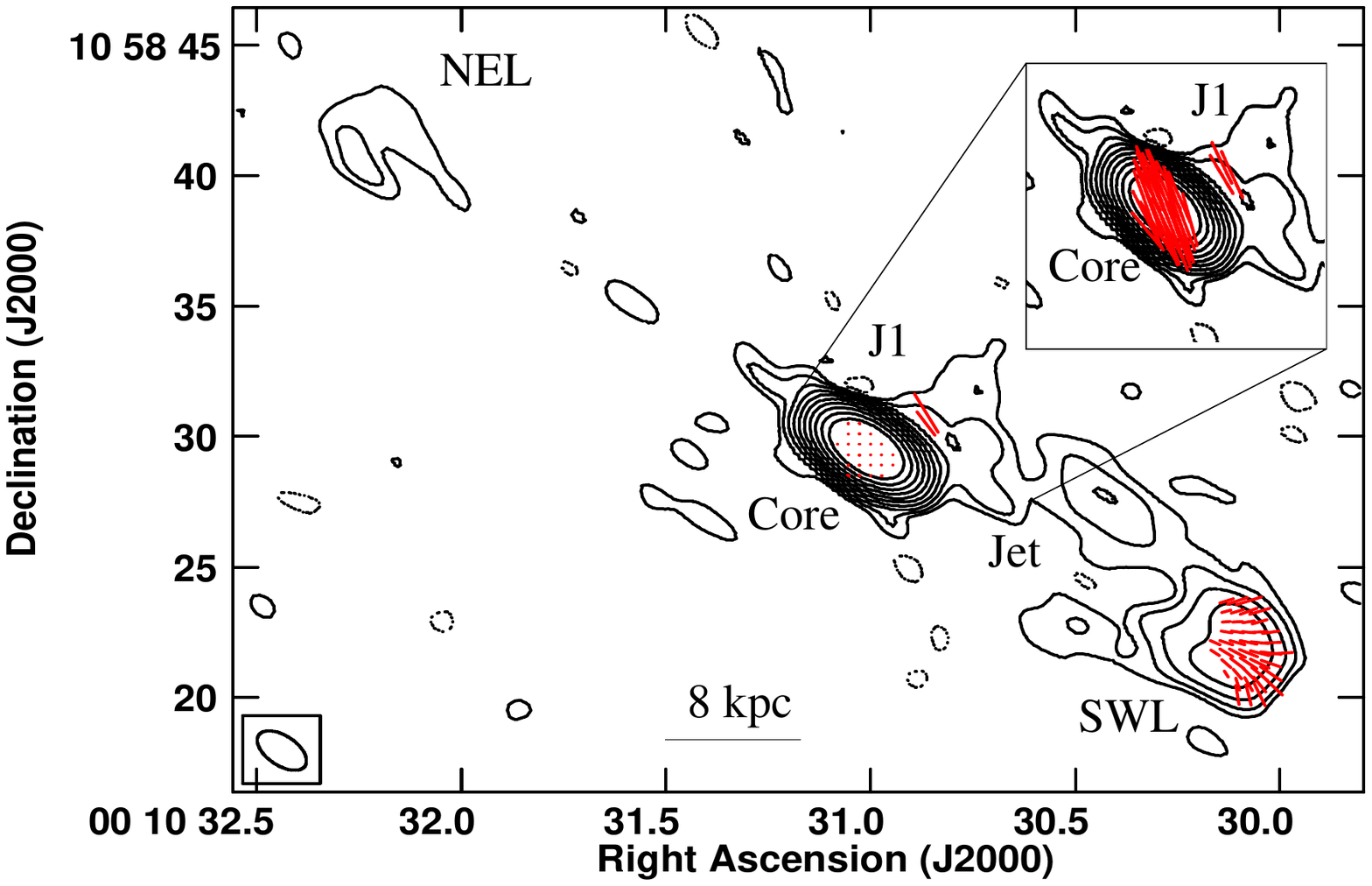}}
\caption{\small VLA 5 GHz total intensity contour image with electric fractional polarization vectors superimposed in red. The inset on the top right presents the VLA 5~GHz total intensity contour image of the radio core with electric polarization vectors superimposed in red. The peak contour flux is 124 mJy beam$^{-1}$ and the contour levels are 0.05 $\times$ (-1, 1, 2, 4, 8, 16, 32, 64, 128, 256, 512) mJy beam$^{-1}$. $1\arcsec$ length of the vector corresponds to 25\% fractional polarization in the main panel and 0.031 mJy beam$^{-1}$ polarized intensity in the inset.}
\label{fig:3}
\end{figure*}

\subsection{uGMRT-VLA RM estimation}
\label{sec:RM_estimation}
The Stokes Q and U images for the VLA 5~GHz data were created from two sub-bands with a bandwidth of 1024 MHz each, obtained from the original data having 16 spectral windows (spw) with a bandwidth of 128 MHz each, using the task $\tt{TCLEAN}$ in $\tt{CASA}$. These were combined to create the polarization angle ($\tt{PANG}$) images using the $\tt{AIPS}$ task $\tt{COMB}$. The mean polarization angle of the ``core'' was estimated from the VLA sub-band $\tt{PANG}$ images as well as the VLA 34 GHz $\tt{PANG}$ image using the task $\tt{IMSTAT}$ in $\tt{AIPS}$. The rotation measure (RM) of the core was estimated by fitting the relation: $\chi$ = $\chi_{\circ}$\,+\,RM\,$\lambda^2$ ($\chi_{\circ}$ and $\chi$ being the intrinsic and observed polarization angles respectively) to the core polarization angles at the 3 frequencies: 5 GHz, 6 GHz and 34 GHz (the former two being the central frequencies of the VLA sub-band datasets), using $\tt{PYTHON}$ fitting procedure - $\tt{CURVE\_FIT}$. The uGMRT core polarization angle was excluded from this analysis since the core polarization structure in the uGMRT image was not clearly resolved from the jet, unlike the VLA images. We note that the calibrated data was not divided into four sub-bands since the signal-to-noise ratio in the individual sub-bands was too low for a reliable RM estimation.

In order to estimate the RM for the south-western (SW) lobe, we created another set of $\tt{PANG}$ images from the VLA sub-band datasets using the Stokes Q and U images that were convolved with an identical circular beam corresponding to the uGMRT resolution (4$\arcsec~\times$ 4$\arcsec$). We also created the uGMRT $\tt{PANG}$ image from the Stokes Q and U images convolved with this beam. The RM for the lobe was estimated by fitting the $\chi-\lambda^2$ relation to the mean polarization angles of the lobe at the 3 frequencies: 685~MHz, 5~GHz and 6~GHz using the $\tt{CURVE\_FIT}$ procedure.

\subsection{uGMRT-VLA Spectral Index Image}
\label{sec:VLA-uGMRT_alpha}
Stokes I images of VLA 1.5 GHz and uGMRT 685 MHz data, convolved with an identical circular beam (6.5$\arcsec~\times$ 6.5$\arcsec$; which is the beam minor of the {\it uv} -tapered uGMRT image; see
Section~\ref{sec:morphology} ahead), were created using the task $\tt{TCLEAN}$ in $\tt{CASA}$. These images were made spatially coincident using the task $\tt{OGEOM}$ in $\tt{AIPS}$, and combined using $\tt{opcode=SPIX}$ in $\tt{AIPS}$ task $\tt{COMB}$. A `two-point' spectral index image was created using the relation: $\alpha$ = log ($\mathrm S_1$/$\mathrm S_2$)/log($\nu_1/\nu_2$) where $\mathrm S_1$ and $\mathrm S_2$ are flux densities at frequencies $\nu_1$ and $\nu_2$ (here $\nu_1$=685 MHz and $\nu_2$=1.5 GHz).  Regions with flux densities below 3 times the {\it rms} noise were blanked while using $\tt{COMB}$.

\section{Results} 
\label{sec:results}
\subsection{Radio Morphology}
\label{sec:morphology}
A triple radio source with a ``hotspot"-core-``hotspot" structure and a curved jet, along with a misaligned lobe, are observed in III~Zw~2. The jet in III~Zw~2 does not terminate in compact hotspots similar to those observed in FRII radio galaxies but rather in ``bow-shock-like'' radio features (see Section~\ref{sec:discussion} ahead). Figure~\ref{fig:1} presents the {\it uv}-tapered uGMRT 685 MHz total intensity contour image of III~Zw~2 superimposed over {\it Sloan Digital Sky Survey} \citep[SDSS;][]{Ahumada20} {\it g}-band optical image, with an inset on the top right presenting the VLA 5 GHz total intensity contour image superimposed over the HST WFC3 F547M optical image of the SW jet and lobe displaying the bow-shock-like morphology of the ``hotspot'' region. The final resolution of the {\it uv}-tapered uGMRT image is 7.0$\arcsec~\times$ 6.5$\arcsec$ with a PA = 88$\degr$. The {\it uv}-tapering was carried out to detect the diffuse lobe emission, especially in the misaligned lobe, better. 

As evident from the figure, the host galaxy is an interacting galaxy and part of an interconnected triplet of compact galaxies. While we cannot rule out contributions to the radio flux density from the companion galaxies seen in projection along the misaligned lobe, we find that the brightest part of diffuse radio emission in the misaligned lobe is not spatially coincident with the second largest of the companion galaxies to the south. Moreover, both the radio flux density and the spectral index of the misaligned lobe provide no clear indications that it has additional contribution from companion galaxies.

\begin{figure}
\includegraphics[width=10cm,trim=20 0 0 350]{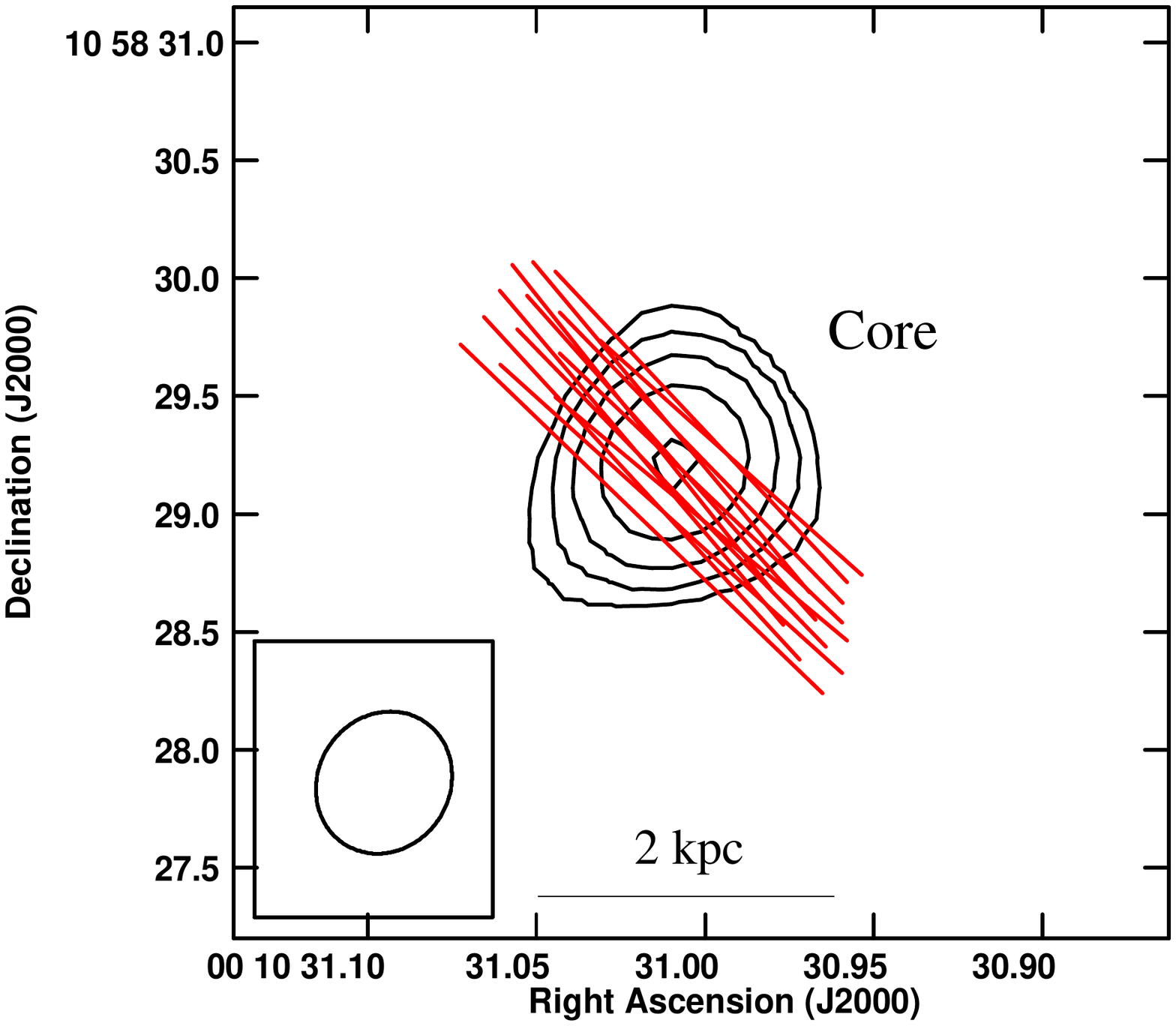}
\caption{\small VLA 34 GHz total intensity contour image with electric fractional polarization vectors superimposed in red. The peak contour flux is 528 mJy beam$^{-1}$ and the contour levels are 30 $\times$ (-1, 1, 2, 4, 8, 16, 32, 64, 128, 256, 512) mJy beam$^{-1}$. $1.5\arcsec$ length of the vector corresponds to 0.3\% fractional polarization.}
\label{fig:4}
\end{figure}

Figures~\ref{fig:2}, \ref{fig:3} and \ref{fig:4} present total intensity contour images of III~Zw~2 with electric fractional polarization vectors superimposed in red using uGMRT at 685 MHz, VLA at 5 GHz in B-array and 34 GHz in C-array respectively. The left panel of Figure~\ref{fig:5} presents a zoom-out of the inset of Figure~\ref{fig:1}, showing the full radio structure of III Zw 2, including the emission from the north-eastern (NE) lobe. From the figure, it appears that the tidal arm towards the north in the optical image may have interacted with the radio jet, but failed to disrupt it. The white curve superimposed over the SW lobe represents the bow-shock-like feature at the jet termination point.

Table ~\ref{tab:Table2} provides the values of total flux density, polarized flux density, fractional polarization and polarization angle for different radio components of III Zw 2 from different projects. The errors in the total flux density of the cores were estimated from the calibration uncertainties
(assuming $\sim$10\%) and {\it rms} noise of the images. Flux density values for compact components like the core were estimated using the Gaussian-fitting $\tt{AIPS}$ task $\tt{JMFIT}$, whereas for extended emission, the $\tt{AIPS}$ tasks $\tt{TVWIN}$ and $\tt{IMSTAT}$ were used. The {\it rms} noise values were also obtained using $\tt{TVWIN}$ and $\tt{IMSTAT}$. The polarized flux density, fractional polarization and polarization angle values were estimated as the mean values for different components from the polarized intensity, fractional polarization and polarization angle images respectively, using the task $\tt{IMSTAT}$ or $\tt{TVSTAT}$ in $\tt{AIPS}$. The respective errors were estimated from the same regions of the corresponding noise images. All the lengths and widths used in the calculations ahead, were measured using the task $\tt{TVDIST}$ in $\tt{AIPS}$.

The distance from the core to ``hotspot'' in the SW lobe is $\sim$ 24~kpc and to the ``hotspot'' in the NE lobe is $\sim$ 33~kpc. The jet is one-sided which could be explained as a result of the Doppler boosting and dimming effects; the brighter SW jet is the approaching one. Also, the jets are asymmetric in length, which could in principle arise from an asymmetry in the environment around the lobes. The medium to the south-west of III~Zw~2 is denser than the medium to the north-east; the closest neighbouring galaxy to the south-west of III~Zw~2 is only $\sim$30 arcsec away \citep[see Figure~\ref{fig:1}; see also][]{Brunthaler05}. We also detect prominently a diffuse lobe to the south-east that is misaligned with the primary lobes (see Figures~\ref{fig:1} and ~\ref{fig:2}). In the 34~GHz VLA image, we detect only the radio core.

The misaligned lobe in III~Zw~2 was briefly discussed by \citet{Silpa20}, where we had presented a shorter exposure uGMRT image at 685~MHz, obtained on December 7, 2018. In that work we had obtained an integrated flux density in the radio core of $\sim$60 mJy and in the current work, it has dropped to $\sim$40 mJy. The $\approx$30\% drop in the core flux density lies within the known observed range in the literature. The integrated flux density of the SW and NE lobes agree within errors between the two epochs. The spectral index image of III~Zw~2 created using our new uGMRT data and similar resolution 1.5 GHz VLA data, indicates a mean spectral index of $-0.7$ and $-0.6$ in the SW and misaligned lobes, respectively (Figure~\ref{fig:7}). 

The spectral index image shows a flat spectrum region in the misaligned lobe. This however is not spatially coincident with the southernmost companion galaxy, with the offset being $-6.4$~arcsec in the north and $+1.5$~arcsec in the east. We also find an ultra-steep ($\alpha<-$1) region to the south of the SW primary lobe. This may be plasma left behind by the jet head that has subsequently aged. Some of this lobe material may also be pushed towards a region of lower ambient density by the bouyant forces from the medium surrounding the SW lobe, giving rise to the misaligned lobe (see Sections~\ref{sec:results_lobe-jet} and ~\ref{sec:discussion} ahead).

\subsection{Polarization Structures \& Energetics}
\label{sec:poln_energetics}
\subsubsection{The Radio Core}
\label{sec:results_core}
The radio core is polarized in the uGMRT 685~MHz image as well as in the 5 and 34~GHz VLA images. For an optically thick core, the inferred B-fields are roughly transverse to the local jet direction and also to the direction of the VLBA jet \citep[P.A for the VLBA jet $\sim74\degr$;][]{Brunthaler00,Brunthaler05}. Using the procedure discussed in Section~\ref{sec:RM_estimation}, we estimated the RM for the core to be $-121\pm15$~rad~m$^{-2}$. We use the following relation to estimate the electron number density (n$_e$) of the Faraday rotating medium that produces an RM of 121 rad~m$^{-2}$ for the core: RM = $\int$ 812\,n$_e$\,B$_{||}$\,dl~rad~m$^{-2}$, where n$_e$ is in cm$^{-3}$, B$_{||}$ is the line-of-sight component of the B-field in milliGauss (mG) and dl is the path length of the Faraday screen in parsecs (pc). Assuming a path length of $\sim$2~kpc, which is the size of the radio core in the 34 GHz VLA image, and B$_{||}$ as the equipartition B-field of 0.14~mG, we obtain an n$_e$ of $\sim5\times10^{-4}$ cm$^{-3}$ which typically corresponds to the number density of the hot ionized medium in the ISM \citep{Spitzer56}. The equipartition B-field is estimated using ``minimum energy'' arguments and equations (1) to (5) from \citet{O'DeaOwen87}, assuming a proton-to-electron energy ratio and a volume filling factor of unity. The radio spectrum is assumed to extend from 10~MHz to 100~GHz. The integrated flux density of the core is estimated from the 34 GHz VLA image and its mean spectral index value (0.8) is estimated using the integrated flux densities at 5 GHz and 34 GHz (see Table~\ref{tab:Table2}). A cylindrical volume with dimensions ($\sim$ 2 kpc, 1 kpc) corresponding to the 34 GHz core is used.

\subsubsection{The Radio Lobes \& Jet}
\label{sec:results_lobe-jet}
The SW lobe is polarized in both uGMRT 685 MHz and VLA 5 GHz images and the jet is polarized in the uGMRT image. Assuming optically thin plasma in the jet and lobes (in which case the B-fields are perpendicular to the $\chi-$vectors), both images reveal that the inferred B-fields are aligned with the edges of the SW lobe and also follow the jet bend at termination. The presence of the Laing-Garrington effect \citep{Garrington88,Laing88} in the lobes of this quasar is also revealed, which favours the SW lobe as the approaching jet side. The uGMRT image reveals B-field to be largely transverse to the jet direction. A jet region close to the core (annotated as J1) in the VLA 5~GHz image is highly polarized, exhibiting inferred B-fields aligned with the local jet direction.

The RM for the SW lobe was estimated as $-$4$\pm$1~rad~m$^{-2}$ using the procedure discussed in Section~\ref{sec:RM_estimation}. Using this value of RM, a path length of $\sim$9 kpc which is the size of the SW lobe estimated from the VLA 5~GHz image, and B$_{||}$ as the equipartition B-field of $\sim7~\mu$G, we obtain an n$_e$ of $\sim8\times10^{-5}$ cm$^{-3}$, which corresponds to the typical number density of the inter-galactic medium \citep[e.g.,][]{ThomsonNelson82}. For the minimum energy calculation, the integrated flux density of the SW lobe is estimated from the VLA 5~GHz image and its mean spectral index value ($-$0.9) is estimated using the integrated flux densities at 685~MHz and 5~GHz (see Table~\ref{tab:Table2}). A cylindrical volume with length and radius of the SW lobe ($\sim$ 9 kpc, 4 kpc) estimated from the 5 GHz image is used.

The equipartition estimates using uGMRT 685 MHz and VLA 1.5 GHz data for both the SW and misaligned lobes are provided in Table~\ref{tab:Table3}. The electron lifetimes are estimated using the relation given in \citet{vanderLaanPerola69}, assuming the break frequency to be 1.5 GHz. The integrated flux densities of the lobes are estimated from the 1.5~GHz VLA image. The mean spectral indices of the lobes are obtained from the 685 MHz$-$1.5 GHz spectral index image presented in Figure~\ref{fig:7}. A cylindrical volume with length and width of the lobes estimated from the 1.5 GHz image is used. 

\begin{figure*}
\centering{
\includegraphics[width=13cm,trim=0 0 70 400]{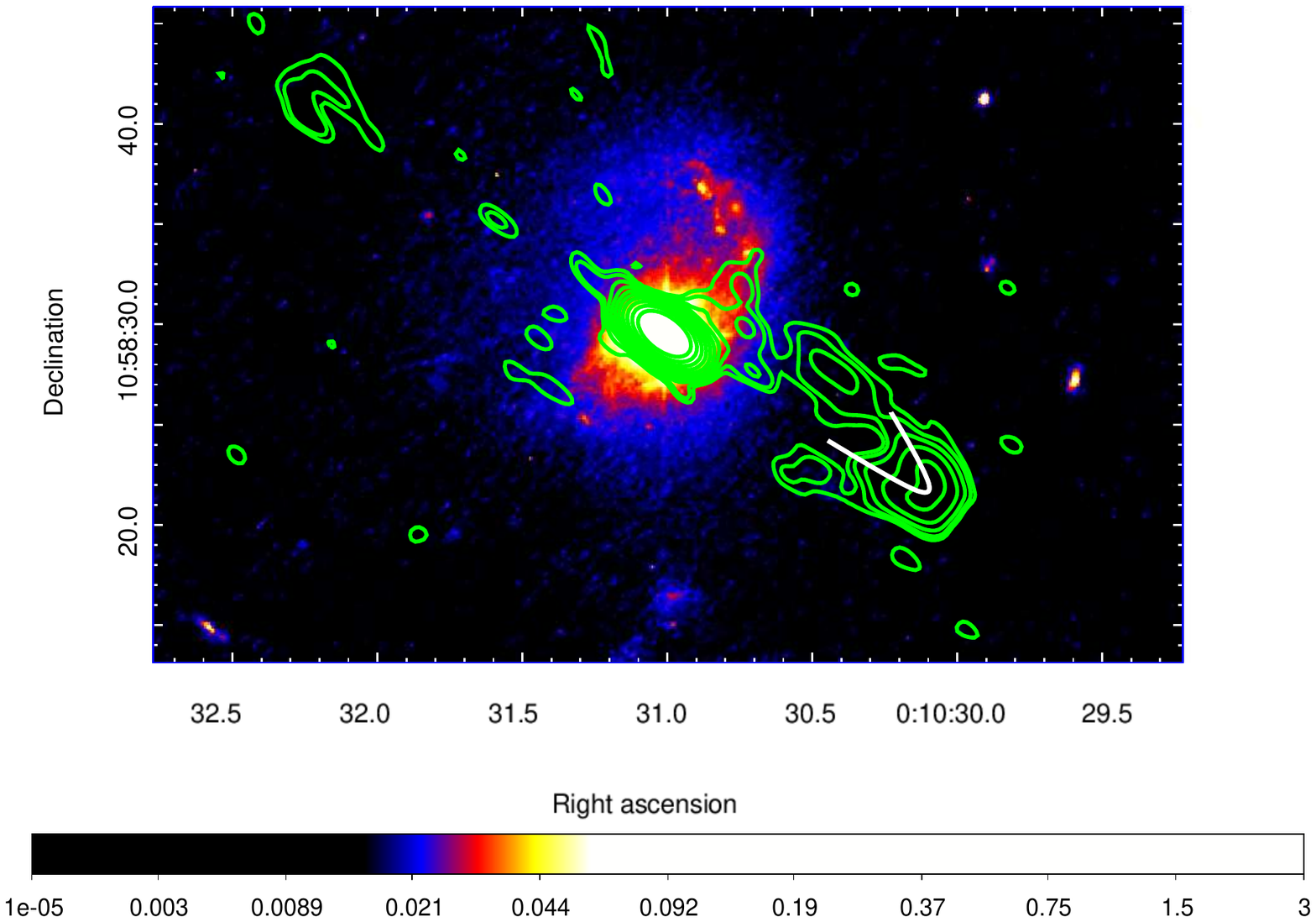}}
\caption{\small VLA 5 GHz total intensity contour image superimposed over the HST WFC3 F547M optical image showing the tidal arm. The contour levels are (0.05, 0.08, 0.14, 0.28, 0.57, 1.21, 2.58, 5.53, 11.88, 25.57) mJy beam$^{-1}$. The color scale extends from $1.0\times10^{-5}$ to $3.0~\times$ ($4.57\times10^{-7}) = 4.6\times10^{-12}$ Jy to 1.4 $\mu$Jy. The white curve represents the bow-shock-like feature at the jet termination point.}
\label{fig:5}
\end{figure*}

\begin{figure}
\centering{
\includegraphics[width=9cm,trim=70 100 0 110]{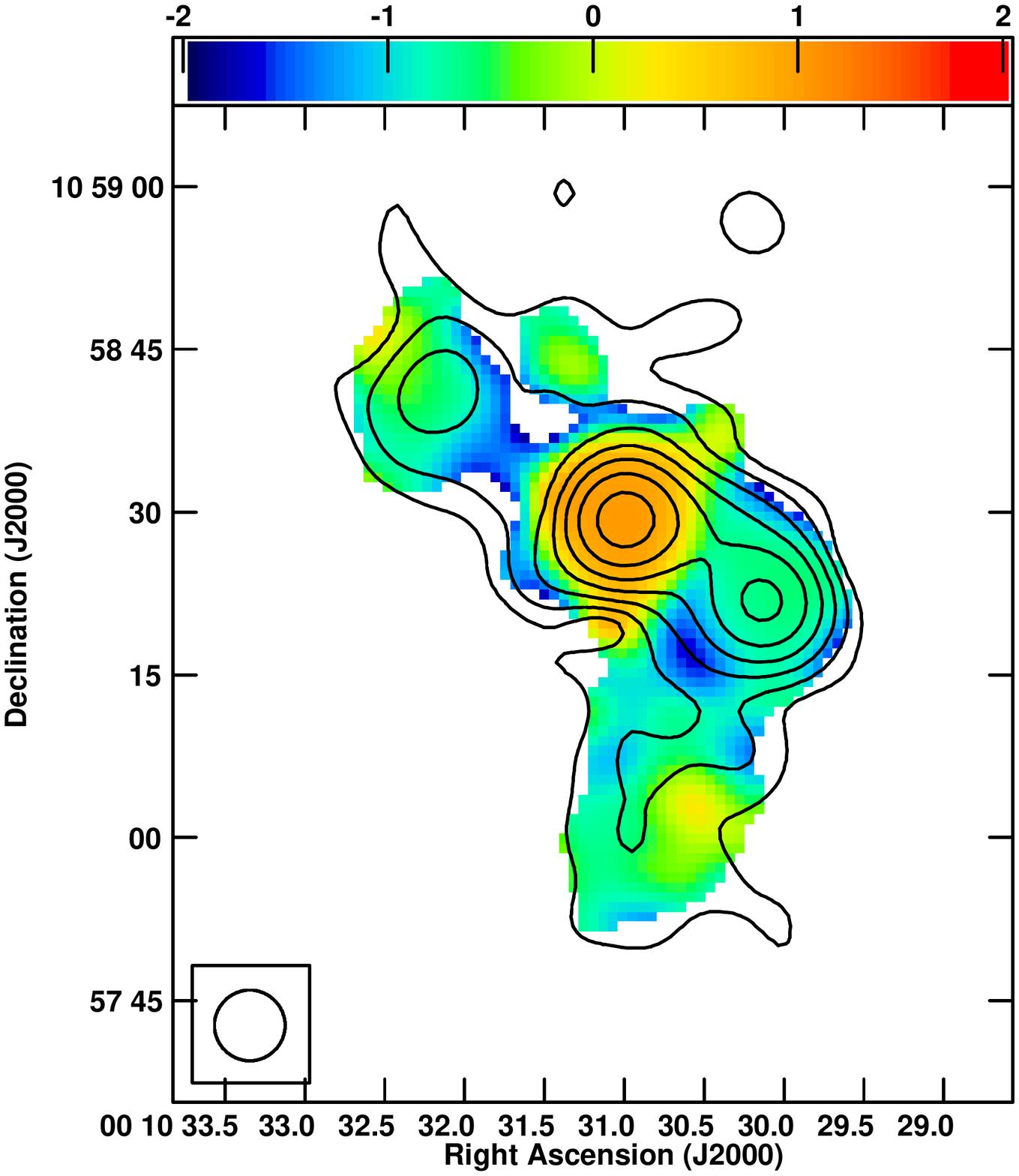}}
\caption{\small 685 MHz uGMRT total intensity contours superimposed over 685 MHz $-$ 1.5 GHz spectral index image in color created using 685 MHz uGMRT and 1.5 GHz VLA data. The peak contour flux is 37 mJy beam$^{-1}$ and the contour levels are 0.4 $\times$ (-1, 1, 2, 4, 8, 16, 32, 64, 128, 256, 512) mJy beam$^{-1}$. The color scale ranges from $-$2 to 2.}
\label{fig:7}
\end{figure}

\begin{table*}
\begin{center}
\caption{Equipartition estimates for the radio lobes at 1.5 GHz}
\label{tab:Table3}
\begin{tabular}{ccccccccc}
\hline
Component & L$_\mathrm{rad}$ & B$_\mathrm{min}$ & E$_\mathrm{min}$ & P$_\mathrm{min}$ &
E$_\mathrm{tot}$ & E$_\mathrm{den}$ & e$^{-}$ lifetime & $\rho_\mathrm{a}$ \\
& (erg~s$^{-1}$) & (Gauss) & (erg) & (dynes~cm$^{-2}$) &
(erg) & (erg~cm$^{-3}$) & (Myr) & (gm cm$^{-3}$)\\
\hline
SW lobe & 3$\times$10$^{40}$ & 4$\times$10$^{-6}$ & 2$\times$10$^{56}$ & 1$\times$10$^{-12}$ & 3$\times$10$^{56}$ & 2$\times$10$^{-12}$ & 34 &
2$\times$10$^{-30}$\\
Misaligned lobe & 2$\times$10$^{40}$ & 2$\times$10$^{-6}$ & 4$\times$10$^{56}$ & 2$\times$10$^{-13}$ & 4$\times$10$^{56}$ & 5$\times$10$^{-13}$ & 32 &
2$\times$10$^{-31}$\\
\hline
\end{tabular}

Column 1: radio source component. Column 2: total radio luminosity. Column 3: B-field at minimum pressure. Column 4: particle energy (electrons and protons) at minimum pressure. Column 5: minimum pressure.  Column 6: total energy  in  particles and fields. Column 7: total energy density. Column 8: lifetime of electrons undergoing both radiative synchrotron and inverse Compton losses on CMB photons. Column 9: density of the ambient medium around the lobes.
\end{center}
\end{table*}

The propagation of the head of a radio lobe through an ambient medium is governed by a balance between the thrust of the jet and the ram pressure exerted by the medium \citep{Schoenmakers00}. The density of the environment around a radio lobe ($\rho_a$) could be estimated as:
$\rho_a$=$\pi_j$/(A$_h$v$_h^2$), where $\pi_j$ is the thrust of the jet spread over a cross-sectional area A$_h$ of the bow-shock at the end of the jet and v$_h$ is the advance speed of the head of the lobe. $\pi_j$ could be written as Q$_j$/v$_j$ where Q$_j$ is the jet power, i.e. the amount of energy delivered by the jet per unit time and v$_j$ is the velocity of the jet material, which is assumed to be c, the velocity of light. v$_h$ could be written as l$_h$/t$_h$ where l$_h$ is the length of the lobe and t$_h$ is the age of the lobe. 

The jet power could be estimated as the total energy of the lobe (from Column 6 of Table~\ref{tab:Table3}) divided by the age of the lobe. We find that both SW and misaligned lobes have comparable ages and jet powers (see Table~\ref{tab:Table3}). A$_h$ is proportional to R$_h^{-2}$, where R$_h$ is the radius of the impact area. Considering half-width of the lobes close to their ends as R$_h$, we find that R$_h$ is $\sim$9 kpc for both the lobes. Also, l$_h$ is $\sim$15 kpc for the SW lobe and $\sim$50~kpc for the misaligned lobe. Higher l$_h$ implies lower $\rho_a$ for the misaligned lobe as compared to the SW lobe.

\section{Discussion}
We discuss below the radio and polarization features in order to explain the nature of the AGN in III~Zw~2.

\subsection{Radio Features}
\label{sec:discussion}

The curved kpc-scale jet terminates in a bow-shock-like region in the SW lobe (see Figure~\ref{fig:5}). A similar structure is also observed at the end of the eastern lobe. A compact terminal hotspot is not observed in III~Zw~2. The inferred B-fields are aligned with the lobe edges. The shape and magnetic field structure resemble that observed at the end of the eastern jet in the episodic radio galaxy, Pictor A \citep{Perley97,Saxton02,Tingay08}, but without a preceding hotspot in III~Zw~2. In fact the shape and magnetic field structures observed at the lobe edge in III~Zw~2 are similar to what is observed at the lobe edges of Seyfert galaxies  \citep{Kharb06,Sebastian19a,Sebastian20}. 

This radio morphology is also observed in Fig. 1e in the restarted jet simulation work of \citet{ClarkeBurns91}. These authors find that when a restarted jet is launched into a region of thermalized plasma from the previous episode of jet activity, and therefore is overdense compared to the old jet material through which it passes, a protrusion is seen at the leading edge of the lobe (their Fig. 1e), similar to the feature observed in III~Zw~2. The morphology of the SW lobe in III~Zw~2 may therefore be indicative of restarted jet activity, which is also supported by the evidence of intermittent jet activity on pc-scales \citep{Brunthaler03, Brunthaler05, Li10}. It is possible that the intermittency has a shorter duty cycle compared to RL AGN. The terminal lobe region could also indicate a sharp jet bend which may have resulted when the jet encountered a density inhomogeneity in the intergalactic medium. However, this cannot explain why a similar structure is observed on the counter-jet side.

Similar spectral indices and electron lifetimes for both SW and misaligned lobes (see Figure~\ref{fig:7} and Table~\ref{tab:Table3}) suggest that the misaligned emission is not due to a previous AGN activity episode. It appears that the outflow, which is likely a combination of a jet and a wind (see Section~\ref{sec:4.2} ahead), moves away from the jet termination region into a lower density region, that is, in the direction of least resistance. This would be consistent with our finding that $\rho_a$ for the misaligned lobe is lower than for the SW lobe (see Section~\ref{sec:results_lobe-jet}).

The primary and the misaligned radio lobes of III~Zw~2 exhibit a mean spectral index between $-$0.6 and $-$0.7 and no clear spectral steepening with distance from the core. These lobe characteristics are similar to those revealed in the ``sputtering'' AGN, NGC 3998 \citep{Sridhar20}. These results could suggest that the primary lobes are being actively fed by the nucleus by intermittent fuelling and that the plasma in the misaligned lobe is being constantly energized and re-accelerated in the presence of turbulence, instabilities and shocks arising from galactic merging activity. 

\subsection{Inferred B-field Structures and Nature of the Outflow}
\label{sec:4.2}
The transverse B-fields in the core are suggestive of a toroidal B-field component at the base of the outflow, which continues all the way up to the edge of the SW lobe as seen in the uGMRT 685~MHz image. The transverse B-fields in the uGMRT jet could either be interpreted as a series of transverse shocks that amplify and order B-fields by compression as seen in BL Lac jets \citep{Gabuzda94,Lister98}, or could represent a toroidal component of a large-scale helical B-field associated with the jet \citep{Pushkarev17}. The parallel B-fields in the jet component J1 from the VLA 5~GHz image could represent a poloidal component of the helical jet B-field. Overall, the polarization structures observed in III~Zw~2 are consistent with the work of \citet{Begelman84}, where the authors suggest that with increasing distance from the core, the toroidal component of the B-field becomes more dominant and the poloidal component decays faster since the poloidal field varies as r$^{-2}$ and the toroidal field varies as r$^{-1}$v$^{-1}$. 

The depolarization of the jet in the 5 GHz VLA image could be explained as an ``inverse depolarization'' effect. This effect has been previously observed by \citet{Homan02, Hovatta12} in optically thin jet features. This effect could originate from a combination of structural inhomogeneities in the jet B-field that could cancel the polarization along the line of sight and internal Faraday rotation that increases with the square of the wavelength \citep{Sokoloff98, Homan12, Homan14}. Such structural inhomogeneities could naturally arise in helical B-fields and random B-fields that are tangled over long length scales. Internal Faraday rotation could align the polarization from the far side of the jet with that from the near side of the jet. This reduces the cancellation between the vectors and increases the net polarization at longer wavelengths.

However, the works of \citet{Mehdipour19} and \citet{Miller12} suggest that toroidal B-fields could be associated with magnetically-driven AGN winds and poloidal B-fields could be associated with jets. Therefore, one cannot rule out the possibility of a magnetized accretion disk wind or the outer layers of a broadened jet (like a jet sheath) threaded by toroidal B-fields, on larger spatial
scales than sampled by the VLA observations and poloidal B-fields threading the spine of the jet. The inherent complexities in the models, data and numerical simulations make it difficult to differentiate between MHD winds and the outer layers of a broadened jet. A co-axial jet and wind outflow model where the jet is embedded inside the wind, would appear co-spatial in projection. Moreover, such a model would be similar to that revealed in the protostellar system HH 212 comprising of a jet and MHD disk wind \citep{Lee21}. 

In order to confirm if the model of \citet{Miller12} and \citet{Mehdipour19} is valid for III~Zw~2, one would require highly sensitive observations at higher frequency and higher resolution that can substantially sample poloidal B-fields in the spine of the jet (for which we already have a marginal detection), in conjunction with emission-line data that can provide complete information on the multi-phase outflow.

By modelling the X-ray reflection spectrum of III~Zw~2 from joint XMM-Newton and NuSTAR observations, \citet{Chamani20} find that the spin parameter for III~Zw~2 is $\geq$0.98. A positive spin parameter suggests a prograde rotation i.e. a co-rotation of the black hole and the accretion disk \citep{Reynolds19}; the high value suggests that the black hole is spinning rapidly. Using an AGN subsample from \citet{Mehdipour19} that included III~Zw~2, \citet{Garofalo19} find that high spinning black hole co-rotating with a radiatively efficient thin accretion disk could give rise to a strong wind and a weak jet. Therefore, we infer that a radiatively efficient thin accretion disc may be present in III Zw 2 at the current epoch.

\subsection{Implications for the Radio Intermediate Nature}
\label{sec:radio-intermediateness}
\citet{Lohfink13} suggest disc truncation and refilling as an explanation for the intermittent jet cycle in 3C120. Moreover, hard and soft spectral states of BHs are often associated with advection-dominated accretion flow \citep[ADAF;][]{NarayanYi94} and thin accretion disk \citep{ShakuraSunyaev73} respectively \citep{Esin97, Kording06}. These could indicate that intermittent jet production is associated with changing spectral states of the accretion disk in microquasars and quasars \citep{Nipoti05}. A soft spectral state is characterized by a standard thin accretion disk that extends to the last stable orbit. Since the magnetic flux accumulation close to the BH in thin accretion disks is not efficient enough for launching of powerful radio jets \citep{SikoraBegelman13}, this state would produce suppressed radio jets and strong winds. Eventually, the inner regions of the standard accretion disk gets destroyed by some instability and replaced by a radiatively inefficient accretion flow. This hard state is accompanied by efficient magnetic flux accumulation close to the BH, giving rise to powerful and steady radio jets. After this the disk refills again with radiatively efficient accretion flow and the spectral state changes to soft state accompanied by quenched radio jets and this cycle repeats. 

Overall, as and when the inner accretion disk region destabilizes, a jet component is ejected. The inner disk region in regular RL and RQ AGN destabilizes on much longer timescales, typically of the order of $10^6-10^8$~yrs \citep{AlexanderLeahy87, TuckerDavid97, EnsslinGopal-Krishna01, McNamara05, Shabala08}, than the AGN showing intermittent/sputtering jet activity on a much shorter, decadal timescales. This may suggest that the RQ AGN are the ones in the soft state, hosting strong winds and suppressed jets, and RL AGN are the ones in the hard state, launching powerful jets, at the time of observations. 

The outflow in III~Zw~2 appears to be a combination of an AGN jet embedded inside a broader wind.
In the framework of changing spectral states of the accretion disk, the MHD wind may be launched from the outer regions of the accretion disk whereas the jet may be launched from the inner regions of the disk. Sputtering/intermittent jet activity, typically on decadal timescales \citep{Brunthaler05, Li10}, may suggest that the time scales of transition between the hard and soft states is so rapid that one may always find a combination of an AGN jet of moderate radio luminosity and an accretion disk wind. 

The interaction of the jet with the wind plasma could possibly disrupt the jet or, much of the mass and energy may be carried away by the wind, rather than the jet. This could explain why the jets in RI AGN are low-powered and small-scaled, as compared to the RL AGN \citep[see also][]{Chamani21}. Moreover, the indication of the restarted jet activity and the associated bow-shock-like morphology in the lobes of III~Zw~2 (discussed in Section~\ref{sec:discussion}) may arise because of the launching of the jet during the current epoch of the hard state into a region of thermalized wind plasma released from the previous epoch of the soft state. The wind component may also be the outer layers of a broadened jet (like a jet sheath) from a previous epoch, or an accretion disc wind or a combination of both. Overall, the close interplay between the jet and the wind may in turn explain the RI nature of III~Zw~2. Moreover, our results are consistent with the idea that radio-loudness is a function of the epoch at which the source is observed \citep[e.g.,][]{Nipoti05, K-B20, Nyland20, Das21}. 

\section{Conclusions} 
\label{sec:conclusions}
We have presented the results from our multi-frequency radio polarization study of the radio-intermediate quasar, III~Zw~2, with the uGMRT and the VLA.
\begin{enumerate}
\item
The inferred B-fields appear to be toroidal at the base of the outflow, which further continues all the way upto the edge of the SW lobe as seen in the uGMRT 685~MHz image. The transverse B-fields in the uGMRT jet may either indicate B-field amplification due to shock compression inside the jet or the toroidal component of a large-scale helical B-field in the jet, or the toroidal B-fields threading an AGN wind that is sampled better by the lower frequency and poorer resolution of the uGMRT compared to the higher frequency and higher resolution of the VLA. The VLA 5~GHz image reveals tentative signatures of a poloidal B-field component in the spine of the jet. Overall, the outflow in III~Zw~2 may be a combination of a jet embedded inside a wind, where the wind can either be a magnetized accretion disk wind or the outer layers of a broadened jet (like a jet sheath) or a combination of both.

\item 
The bow-shock-like morphology of the jet terminal regions are consistent with restarted jet activity in III~Zw~2, which is in turn consistent with its ``sputtering'' nature. 

\item 
Diffuse lobe emission that is misaligned with the primary lobes is detected. The spectral indices and the electron lifetimes in the misaligned lobe are similar to those in the SW lobe, suggesting that the misaligned emission is not due to a previous AGN activity episode. It appears that the SW lobe is embedded in a medium of higher density than the misaligned lobe. Therefore, the buoyant forces applied by the medium around the SW lobe pushes the lobe material laterally to a region of lower density so as to achieve a density equilibrium, which gives rise to the misaligned lobe emission.

\item 
Spectral state changes in the accretion disk and the subsequent episodic/intermittent behaviour of the outflow, along with the close interplay between the AGN jet and wind, could be responsible for the radio-intermediate nature of III~Zw~2. Moreover, based on suggestions of a co-existence of an AGN jet and wind, and the presence of a prograde-rotating high spin BH, we infer that a radiatively efficient thin accretion disk may be present in III~Zw~2 at the current epoch, as proposed in the work of \citet{Garofalo19}.
\end{enumerate}

The current work clearly demonstrates the power of radio polarization studies to understand the origin of radio outflows in systems where multiple emission mechanisms co-exist, such as the RI AGN. The fact that the RI AGN straddle the RL-RQ division, makes them ideal candidates for not only understanding the origin of the RL-RQ dichotomy but also providing insights on jets and winds and their impact on galaxy evolution via feedback processes.

\section*{Acknowledgements}
We thank the referee for their insightful suggestions that have improved this manuscript significantly. LCH was supported by the National Science Foundation of China (11721303, 11991052) and the National Key R\&D Program of China (2016YFA0400702). We thank the staff of the GMRT who have made these observations possible. The GMRT is run by the National Centre for Radio Astrophysics of the Tata Institute of Fundamental Research. We acknowledge the support of the Department of Atomic Energy, Government of India, under the project 12-R\&D-TFR-5.02-0700. The National Radio Astronomy Observatory is a facility of the National Science Foundation operated under cooperative agreement by Associated Universities, Inc.

\section*{Data Availability}
The data underlying this article will be shared on reasonable request to the corresponding author. The VLA data underlying this article can be obtained from the NRAO Science Data Archive (https://archive.nrao.edu/archive/advquery.jsp) using the proposal ids: 20A-182, 17A-027 and 15A-070. The GMRT data underlying this article can be obtained from the GMRT Online Archive (https://naps.ncra.tifr.res.in/goa/data/search) using the proposal id: DDTC130. The HST data underlying this article can be obtained from the MAST HST Archive (https://archive.stsci.edu/hst/search.php) using the proposal id: ICNO01010.



\bibliographystyle{mnras}
\bibliography{ms} 








\bsp	
\label{lastpage}
\end{document}